\documentstyle[prb,aps,twoside,twocolumn,epsf]{revtex}
\begin{document}

\title{\boldmath
Low-temperature anomalous specific heat without
tunneling modes: a simulation for a-Si with voids
}
\author{
 Serge M. Nakhmanson and
 D. A. Drabold\\
 {\small\it
  Department of Physics and Astronomy, 
  Condensed Matter and Surface Science Program,
  Ohio~University,
  Athens, Ohio 45701-2979\\ 
 }
}
\date{\today}
\maketitle

\begin{abstract}
\noindent
Using empirical potential molecular dynamics we compute dynamical matrix
eigenvalues and eigenvectors for a 4096 atom model of amorphous silicon and 
a set of models with voids of different size based on it. This information 
is then employed to study the localization properties of the low-energy
vibrational states,
calculate the specific heat $C(T)$ and examine
the low-temperature properties of our models usually attributed to the
presence of tunneling states in amorphous silicon. The results of our
calculations for $C(T)$ and ``excess specific heat bulge'' in the
\hbox{$C(T)/T^3$} vs.\ $T$ graph for {\it voidless} a-Si appear to be in good 
agreement with experiment; moreover our investigation shows that
the presence of localized low-energy excitations in 
the vibrational spectrum of our
models {\it with voids} strongly
manifests itself as a sharp peak in \hbox{$C(T)/T^3$}
dependence at $T < 3K$. To our knowledge this is the first numerical
simulation that provides adequate agreement with experiment for
the very low-temperature properties of specific heat in disordered
systems {\it within} the limits of harmonic approximation.

\end{abstract}

\section {Introduction}
The main goal of the computational project 
presented below was to construct a set of realistic models for
device quality a-Si (a-Si:H) material, containing
nanovoids in the structure, and explore the vibrational properties
of these models, especially the localization patterns of the
low-energy states that emerge after introducing a void into
silicon continuous random network (CRN). Modern 
computational facilities allow us to
study models of up to several hundreds of atoms with
{\it ab initio} methods and thousands of atoms if we switch to 
empirical techniques, which is especially helpful for realistic large scale
modeling of amorphous materials. We can perform simulated
quenching and annealing for these models, looking for their
most energetically favorable geometry, then we can calculate
the dynamical matrices for our systems and diagonalize them
exactly, receiving their eigenvalues together with the conjugate
eigenvectors. This data gives us the ability to construct the
vibrational density of states (VDOS) for a given model as well
as to look at the localization properties of individual
eigenstates of its dynamical matrix. The calculation of
specific heat can be seen as a natural extension of these 
 techniques because we can do it with a little 
effort once we have obtained the VDOS information for
the model.

In Section II of this paper we present the details of our
model construction scheme for a large (based on 4096
atom model) family of models for a-Si with voids 
and introduce a set of methods we
employ for geometry optimization, dynamical matrix and
specific heat calculations. In Section III 
we discuss the results of our calculations
of vibrational properties and  specific heat for
the models considered and finally in Section IV we summarize our
findings about the influence of localized low-energy
vibrational modes on the thermodynamical properties of
amorphous silicon.

\section{Model construction and computational procedures}

We use the Djordjevic, Thorpe and Wooten 
4096 atom model\cite{WWWupd} for a-Si
(referred to as DTW in what follows) constructed with the Wooten, Winer
and Weaire bond switching algorithm\cite{WWW} as a base for
building a family of models with voids. The length of the side of 
a cubic supercell for this model is approximately 43~\AA.\
As our first step we optimize the geometry of the basic model
which results only in minor network rearrangements; this relaxed
geometry is then used for producing all of the
following models with voids. To cut out a void we pick an
arbitrary atom in the network and 
remove it as well as the consecutive spherical shells of its
neighbors. We find that our results do not depend much on
which atom we select for this procedure.

By applying this procedure we have built three models with voids
of different diameter: a 4091 atom model with a ``small
void'' (only one atom and four of its nearest neighbors removed) --- a void
of approximately 5~\AA\ in diameter, 4069 atom ``medium void''
model with 10~\AA\ void and 4008 atom ``large void'' model
with 15~\AA\ void. We refrained from building models with even
larger voids to prevent possible interaction of a void with its
own ``ghost'' images in the neighboring periodically translated
supercells.

Every model with voids was then quenched to minimize the forces
acting on atoms in the network. The atomic forces 
must be small for the application of the harmonic approximation 
for the total energy of the system, which is required for the
dynamical matrix calculation. After the dynamical matrix
calculation for every model mentioned above,
the eigenvalue and eigenvector data was
used to produce VDOS and inverse participation ratio\cite{ipr} 
(IPR) graphs for the model, calculate its specific heat and
visualize the spatial localization/delocalization characteristics
of some of its vibrational modes (the ones which behavior we 
found most interesting).

For geometry optimization of the models (simulated quenching) 
and dynamical matrix calculation we 
employ a molecular dynamics code ``Estrelle'' developed
by the authors of this paper.  The code is built around an empirical
environment-dependent interatomic potential (EDIP), which
has been recently introduced by Bazant and Kaxiras\cite{EDIP1,EDIP2,EDIP3}.
In general, this potential inherits the well established 
Stillinger-Weber\cite{SW} format for two- and three-body interactions,
but now these interaction terms depend on the local atomic environment
through an effective coordination parameter.
Our testing results for EDIP and its performance in comparison
to our previous {\it ab initio} calculations\cite{ai_voids} 
are described elsewhere\cite{voids_JNCS}.

The force tolerance threshold in simulated quenching mode of
our MD program is set to be $0.01$~eV/\AA\ for all its applications we discuss here.
The dynamical matrix for any given model is computed by displacing
every atom in the cell by $0.03$~\AA\  in three orthogonal directions
and calculating the originating forces on all the atoms in the 
system\cite{DMatrix}. 
For a system of thousands atoms the dynamical matrix is very large which,
under normal circumstances, causes problems in storing it
on disk or in computer memory. Fortunately the dynamical
matrix is also very sparse, because in most cases the displacement
of a single atom generates significant forces only on its
closest neighbors but not in the whole supercell. We extensively exploit 
this localization of dynamical matrices in our calculations, 
discarding terms smaller than 
\hbox{$10^{-4} \hbox{eV} \hbox{\AA}^{-2} {a.u.m.}^{-1}$},
which is a good compromise between accuracy and
compactness of the output.
Once the sparse dynamical matrix for the system is obtained we use a separate
routine to exactly diagonalize the
whole matrix and obtain all of the eigenvalues and eigenvectors. 
Again, for the same reasons as already mentioned above, we do
not write out all of the eigenvectors (however, we do keep all their IPRs)
but rather only those that exhibit properties we look for:
(i) small energy/eigenvalue (less than \hbox{$200~cm^{-1}$}) and (ii)
relatively high IPR, showing that the vibrational mode we are dealing
with is localized.

To create VDOS graphs we employ a Gaus\-si\-an  
representation for \hbox{$\delta(E-E_i)$}, where
\hbox{$E_i, i = 1 \ldots N$} are the eigenvalues of the dynamical
matrix and $N=3N_{atoms}$. The width of broadening is \hbox{$20~cm^{-1}$} for the
full scale graphs and \hbox{$0.1~cm^{-1}$} for the close-ups of the
low-energy region. The vibrational activity colormaps for the 
``low-energy, high IPR'' modes are prepared in
same way that has been already described in Sec.\ II of Ref.~8:
the set of individual atomic IPRs is computed and then every atom
is assigned a certain color according to its displacement from the
equilibrium position.

It is relatively easy to obtain $C(T)$ dependence for the
model if the VDOS information for it is available\cite{Maradudin}:
$$
C(T) = 3R\int_0^{E_{max}}{ \Big( {{E}\over{k_B T}} \Big)^2 
{ {e^{E/k_BT}}\over{\big( e^{E/k_BT} - 1 \big)^2} }\, g(E)\,dE },
$$
where VDOS $g(E)$ is normalized to unity.
Nevertheless one thing should be treated with caution: the model VDOS
one usually has is relevant for a system of {\it finite}
size (i. e. our supercell). Vibrational excitations
with wavelengths longer than the size of the supercell cannot be excited
in this model and are consequently missing in its VDOS data. In 
order to receive the precise values for $C(T)$ one should correct VDOS for the
{\it infinite} size of the system. In our case it is done in the following
fashion: all the delocalized (acoustic) vibrational modes of energy less than 
\hbox{20~$cm^{-1}$} are cut out and substituted by a weak parabolic 
tail $\alpha E^2$ in the routine to compute the VDOS. Parameter $\alpha$
can be obtained from a calculation of the elastic constants of the
model\cite{Feldman1} but in this investigation we use a more
simple approach, fitting $\alpha$ to provide a smooth transition between
the low-energy parabolic tail and the rest of VDOS.

\section {Discussion of results}

\subsection{Vibrational properties and localization}

\begin{figure*}[h]

\epsfxsize=7cm
\epsfysize=10cm
\moveright 0.7cm \vbox{\epsfbox{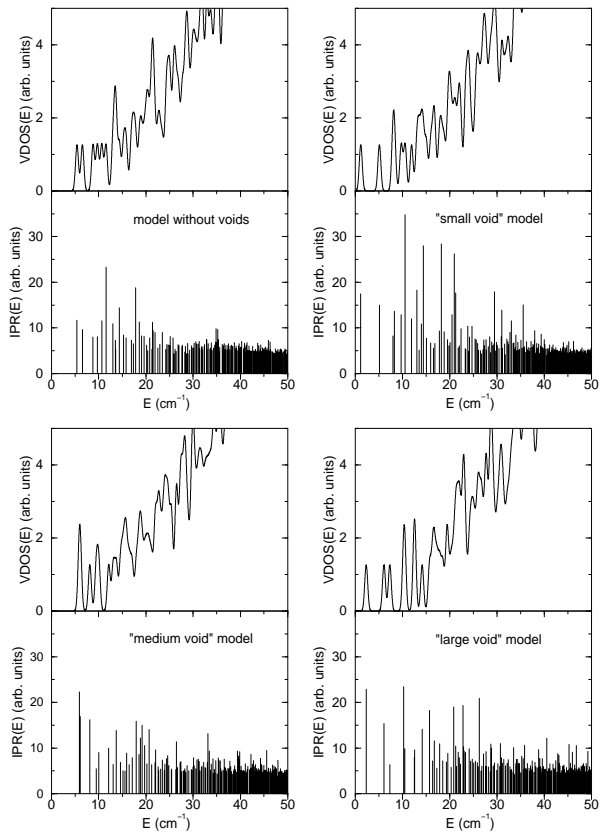}}

\vskip 0.7cm

\caption{
\label{big_four}
Low-energy VDOS and IPR snapshots for 4096 atom DTW model
without voids
(upper left set of panels), 4091 atom ``small void''
model (upper right), 4069 atom ``medium void''
model (lower left) and 4008 atom ``large void''
model (lower right).}
\end{figure*}

We begin this section by presenting the results of our calculations
of the low-energy regions\cite{full_VDOS} for VDOS and IPR for 
all the four models introduced above, shown in Fig.~\ref{big_four}. 
For the sake of simplicity we do not present the colormaps for
vibrational modes for 4096 atom family of models in this 
paper, however a set of colormaps for the most interesting
localized excitations in these models is available for download
over the World Wide Web\cite{download}.

From Fig.~\ref{big_four} we can see that the large models for a-Si
(both, with and without voids) exhibit quite a complicated vibrational
behavior, much more complex than that of smaller 216 atom based
families of models, we have studied before\cite{ai_voids,voids_JNCS}.
The most important difference here is that a-Si model {\it without voids} 
has two localized
low-energy modes that are associated with strained
regions of silicon network: we have checked bond lengths and bond
angles for the atoms in these regions and found that the modes
generally localize on atoms with bond angles deviating from the perfect
tetrahedral angle by more than 30 degrees.

Consequently, now we have two types of {\it phonon traps} in our
models with voids --- the voids themselves and strained regions of the 
network. Keeping this in mind we can attempt to
introduce a rough classification of the localized low-energy modes
according to the type of phonon trap they fall into.
First, we can see a significant number of vibrational modes in our models
with voids
that generally show the same kind of localization properties that we have 
reported earlier\cite{ai_voids}: they are exponentially localized
with the center of localization positioned
to the side of the void. We classify these excitations as void type
modes. Secondly, the modes we might attribute to the {\it strained network
region phonon trap} type
in models with voids exhibit a different kind of behavior in comparison
to the voidless model. These modes do not localize {\it exactly} on the strained
regions in the supercell; instead they form a string extended between
one of these strained regions and the void\cite{JJ}. The
possible explanation of this behavior is that these modes can be regarded as a
superposition of void type modes and the localized excitations in
the model without voids. In our classification we call them mixed type
modes. We have to stress once again that the classification we propose
is only approximate and is based mostly on the colormaps (i.\ e.\ pictures)
we get for our models {\it not} on rigorous mathematical arguments.
We must also add that all the low-energy modes, that appear localized in
our {\it finite} models, will be pseudolocalized in an infinite 
sample\cite{Feldman2}.

\vskip 0.7cm

We must admit that in our current investigation we were not able
to find any simple connection between the size of the void and the energy
and type of resulting localized modes.
Our data shows that for different models with voids modes
of different types dominate in the low-energy region. In the ``small void'' 
model a mode with the
highest IPR at \hbox{10.58~$cm^{-1}$} is of void type, but the succeeding
three modes with high IPR at 14.43, 18.25 and \hbox{20.97~$cm^{-1}$}
are of strongly pronounced mixed type. In the ``medium void'' model,
to the contrary, all three low-energy localized modes at 5.89, 6.12
and \hbox{8.13~$cm^{-1}$} are of mixed type. The mode with strong
void type behavior is also present but it is shifted to 
\hbox{17.97~$cm^{-1}$}. Finally in the ``large void'' model
modes at 2.34 and \hbox{6.10~$cm^{-1}$} are of void type and all the
others, including a strongly localized mode at \hbox{10.28~$cm^{-1}$},
exhibit mixed type behavior. We {\it speculate} that the network strain
and geometrical peculiarities of any given model play a more important role
in shaping the energy and type distribution of its localized vibrational
modes than the actual size of the void --- at least for the models with
voids of comparable sizes, as we have here.

\subsection{Specific heat}

In this section we present our results for the calculations 
of specific heat $C(T)$
for the family of 4096 atom models.
The overall temperature dependence
for specific heat for all of our models is in good agreement with
Dulong and Petit's law at high temperatures and Debye's law at
low temperatures; our calculation also produces
approximately the correct Debye temperature for a-Si.
For the room temperature (300K) we receive
practically the same value for specific heat for all of our models:
\hbox{$19.7~JK^{-1}mol^{-1}$}.

\begin{figure*}[h]

\epsfxsize=7cm
\moveright 0.7cm \vbox{\epsfbox{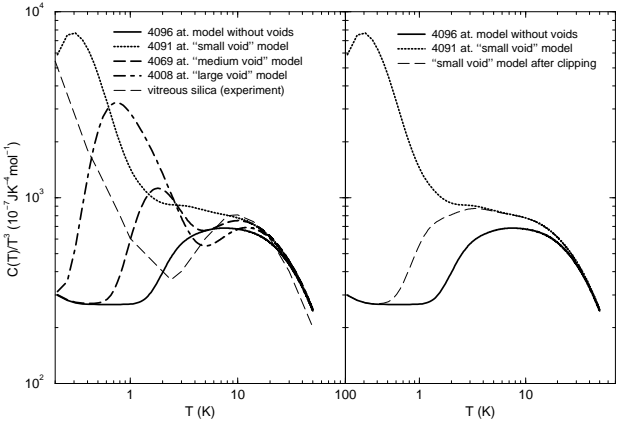}}

\vskip -2.0cm

\caption{ 
\label{C_of_TTT}
Left panel: the low-temperature \hbox{$C(T)/T^3$} dependence for
the DTW models with and without voids. The experimental curve for vitreous silica
is taken from Ref.~18.
Right panel: the curves for 4091 atom ``small void'' model before 
and after clipping
of the lowest energy localized mode eigenvalue. Curve for the
model without voids is also shown for reference.
}
\end{figure*}

In the left panel of  Fig.~\ref{C_of_TTT} the \hbox{$C(T)/T^3$} 
low-temperature 
dependence for our models is presented. The most striking feature
in this graph is the presence of sharp peaks at $T < 3K$ in the
curves for the models {\it containing voids}. The model without voids
{\it does not} have this peak, although it does demonstrate the
presence of the well known excess specific heat bulge,  
the position and height of which are in qualitative
agreement with experiment\cite{Mertig} as well as with recent
computational results of Feldman, Allen and Bickham\cite{Feldman2}.
All of our models with voids also have the excess specific heat bulges
at approximately the same position, but comparing to the
low-temperature peaks their intensities are about an order of
magnitude smaller. We were not able to find any experimental
data for specific heat measurements in a-Si at temperatures below
$2K$, but in order to make some general comparison to experiment
for these new low-temperature features we obtain (which should be
generic for {\it any} disordered system containing voids), we provide 
the experimental curve for vitreous silica\cite{TSG} in our graph.

Unlike the previous void size vs.\ energy situation, we can
find a clear connection between the presence of low-energy localized
modes in vibrational spectrum of the model and the height (or even
absence) of the low-temperature peak in its \hbox{$C(T)/T^3$}
graph. For the ``small void'' model we have a localized mode at
\hbox{1.19~$cm^{-1}$} (here and below, see Fig.~\ref{big_four})
 --- the lowest energy at which we can see localized
excitations in all our models --- and the highest peak in
\hbox{$C(T)/T^3$} dependence. The ``large void'' model has its
lowest energy localized excitation at \hbox{2.34~$cm^{-1}$} and the
peak of smaller height comparing to the previous model. The
``medium void'' model has two localized states but only at
approximately \hbox{6~$cm^{-1}$} and peak that is even less 
pronounced than in case of the first two models. Finally the
model without voids has no localized states with energy lower than
\hbox{11~$cm^{-1}$} and {\it no low-temperature peak whatsoever.}

\vskip -2.0cm

In order to investigate this connection in more detail we have
performed a simple numerical experiment, which results are
shown in 
the right panel of  Fig.~\ref{C_of_TTT}. We have clipped the eigenvalue at
\hbox{1.19$cm^{-1}$} from the eigenvalue set for the ``small
void'' model and recalculated its VDOS and $C(T)$ receiving
no low-temperature peak in \hbox{$C(T)/T^3$} graph, much
like in the situation with the model without voids. In our
opinion these results provide enough evidence to attribute the existence
of low-temperature (\hbox{$T < 3K$}) peak  in \hbox{$C(T)/T^3$} dependence
for the model to the presence of localized low-energy 
(\hbox{$E \sim 1 - 6~cm^{-1}$}) vibrational
excitations --- in our case produced by voids --- in its spectrum.

Finally we must note that the localized vibrational excitations we see,
although having rather low energies, are {\it not} tunneling states\cite{TSG},
that are nonharmonic by nature and can not be obtained in harmonic
approximation calculation. We do not claim that the whole tunneling
states theory is incorrect, we rather propose an alternative mechanism that
explains the same experimental data. It seems that {\it any}
mechanism that creates additional density of vibrational states
(be this tunneling states or low-energy localized ``void'' vibrations in porous
materials) at very low energies will produce the same effect on low-temperature
specific heat behavior. In order to find out which mechanism of the
two mentioned above {\it actually works} in real material, an experimental
investigation of low-temperature thermal properties 
and {\it simultaneously} geometrical quality (i.e. presence of defects,
voids, strained regions) of this material should be carried out. The works
of X. Liu {\it et al.\cite{Liu1,Liu2}} or Coeck and Laermans\cite{Coeck}
for amorphous silicon can be regarded as the closest examples here.

\section {Conclusions}
We have studied vibrational and thermodynamical properties of 4096 atom 
DTW model for amorphous silicon and the family of models with voids 
based on it, employing Bazant-Kaxiras en\-vi\-ron\-ment-de\-pen\-dent 
interatomic potential and empirical MD technique. 
We have found that the models with voids 
posses a complex spectrum of localized low-energy excitations
that can be {\it roughly} divided into two groups --- void and mixed 
type modes --- according to their localization patterns.
Our calculations show that there is no simple connection between
the size of the void and the energy and type of its localized modes. 
It is most probable that not only the size of the void but also its
local geometrical environment as well as strain distribution
in the neighboring regions of the network play a paramount role in
shaping the low-energy vibrational spectrum of the system.
We have constructed specific heat $C(T)$ plots for our models,
that appear to be in good agreement with experiment. We have
also plotted out our models' \hbox{$C(T)/T^3$} dependences for the
low-temperature region, which seem to be in adequate agreement
with experimental and other computational results for 
\hbox{$T > 3K$} (the excess specific heat bulge) and predict new
interesting features, undoubtedly connected with vibrational 
properties of voids present in the system, at lower temperatures. 
 We must stress
that our results are correct for model materials with a {\it uniform}
distribution of voids of {\it one and the same size}, which is of
course impossible to produce in real material. Nevertheless, employing
our model data we can predict that in real material the localized
low-energy vibrational states, connected to voids of different sizes, 
will fill out a band which will alter the parabolic VDOS tail properties
at small energies and consequently manifest itself by changing the
specific heat \hbox{$C(T)/T^3$} dependence.

\section*{Acknowledgments}
This work was supported by NSF
under grant number DMR 96-04921 and DMR 96-18789. We thank Prof.\ Normand Mousseau
for many helpful discussions.

\end{document}